\documentclass[12pt,prd,showpacs,tightenlines,nofootinbib]{revtex4}
\usepackage{bm}
\usepackage{graphicx}
\usepackage{rotating}
\usepackage{epsfig}
\begin{document}

\title{Rare $B\to \pi l\bar l$ and $B\to\rho l \bar l$ decays in the
  relativistic quark model } 

\author{R. N. Faustov}
\author{V. O. Galkin}
\affiliation{Dorodnicyn Computing Centre, Russian Academy of Sciences,
  Vavilov Str. 40, 119333 Moscow, Russia}

\begin{abstract}
The branching fractions of the rare weak $B\to \pi l^+l^-(\nu\bar\nu)$ and $B\to\rho l^+l^-(\nu\bar\nu)$ decays are calculated in the framework of the relativistic quark model based on the quasipotential approach. The form factors parametrizing weak decay matrix elements are explicitly determined in the whole kinematical $q^2$ range without additional assumptions and extrapolations. Relativistic effects are systematically taken into account including recoil effects in meson wave functions and contributions of the intermediate negative-energy states. New experimental data on the differential distributions in the semileptonic heavy-to-light $B\to\pi l\nu_l$ and $B\to\rho l\nu_l$ decays are analyzed in detail. Good agreement of the predictions and data is found. The obtained results for the branching fractions of the rare semileptonic decays are found to be in agreement with other theoretical estimates and recent experimental data available for the $B^+\to\pi^+ \mu^+\mu^-$ decay.    

\end{abstract}

\pacs{ 13.20.He, 12.39.Ki}

\maketitle

\section{Introduction}
\label{sec:int}

Recently significant experimental progress has been achieved in
studying weak heavy-to-light decays of $B$ mesons. For the
semileptonic $B\to\pi l\nu_l$ and $B\to\rho l\nu_l$ decays not only
total decay branching fractions were measured by Belle
and BaBar Collaborations
\cite{Belle1,Belle2,Babar} rather precisely but
also differential distributions in rather narrow $q^2$ bins. Such measurements are very important since they provide the test of the momentum dependence of the weak differential decay branching
fractions and thus significantly constrain theoretical models. It also
allows us to extract  the Cabibbo-Kobayashi-Maskawa (CKM) matrix  element $V_{ub}$ from exclusive decay channels with better
precision and confront it with the value obtained from inclusive
semileptonic decays. Moreover, recently the LHCb Collaboration \cite{lhcbBpi}
reported first observation of the rare $B^+\to\pi^+\mu^+\mu^-$ decay 
with the branching fraction
$Br(B^+\to\pi^+\mu^+\mu^-)=(2.3\pm0.6\pm0.1)\times 10^{-8}$. Such
decays are governed by the flavour changing neutral current and thus
are very sensitive to the contributions of new intermediate particles
and interactions. Therefore the study of rare $B$ decays is important
for constraining the theories which go beyond the standard model. Since
these decays are induced by loop diagrams they are suppressed by at
least three orders of magnitude compared to corresponding
heavy-to-light semileptonic $B$ decays. Observation of the rare
$B\to\pi\mu^+\mu^-$ decay signifies an important progress since this
decay is stronger CKM suppressed compared with better studied $B\to
K^{(*)}\mu^+\mu^-$ decays.  

Theoretical investigation of weak decays requires the determination of the
decay matrix elements of the weak current between meson states. It is
convenient to parametrize these decay matrix elements in terms of the
invariant form factors. The calculation of these form factors demands
application of the nonperturbative methods. Since these decays are
governed by the heavy-to-light quark transitions they have a very
broad kinematical range. Various theoretical approaches were used to
calculate these form factors. However most of the employed methods
allow determination of the momentum transfer dependence of
the form factors only in a rather limited $q^2$ range. For example light
cone QCD sum rules are applicable in the large recoil region
($q^2\approx 0$) while lattice QCD provides results at small recoil
($q^2\approx q^2_{\rm max}$). Therefore some model assumptions and/or
parametrizations should be used to extrapolate the results in the
whole range of $q^2$, thus introducing additional theoretical uncertainties.    

In our previous papers \cite{rarebk,rarebs} we investigated the rare $B\to
K^{(*)}l^+l^-$ and rare $B_s$ decays in the framework of the
relativistic quark model  with the QCD motivated quasipotential of the quark-antiquark interaction. Calculating the decay form factors we systematically 
took into account relativistic effects including transformations of the
meson wave functions from the rest to the moving reference frame and
contributions of the intermediate negative-energy states. All form factors are expressed through the usual overlap integrals of the meson wave functions, which are known from the previous mass spectrum considerations
\cite{mass,lm,hlm}. The important advantage of the
developed method consists in the fact that it provides explicit calculation of
the momentum transfer dependence of the form factors in the whole kinematical range, thus improving the reliability of the obtained results. Here we
apply this approach for studying the rare $B\to\pi(\rho)l\bar l$ decays. Evaluation of such decay branching fractions requires calculation of the matrix
elements of weak vector, axial vector and tensor currents. In
Ref.~\cite{asld} we calculated the form factors parametrizing matrix
elements of the weak vector and axial vector currents for the $B\to\pi$ and
$B\to\rho$ transitions and on this basis studied the corresponding
semileptonic decays $B\to\pi(\rho)l\nu_l$. We first
confront the predictions of our model with new detailed Belle and
BaBar data \cite{Belle1,Belle2,Babar} on heavy-to-light semileptonic
decays. Such comparison provides additional test the $q^2$ dependence of the form factors. Then we apply the model
for the calculation of the tensor form factors. On this basis
the differential distributions and total decay branching fractions as well as the
forward-backward asymmetry and the $\rho$ polarization fractions are
calculated and confronted with available experimental data and other
theoretical predictions.       

\section{Form factors of the weak $B$ meson transitions to $\pi$ and
  $\rho$ mesons in the relativistic quark model}
\label{sec:ffbpi}

The matrix elements of the weak current for the heavy-to-light $b\to q$ ($q=u,d$) weak transitions  between the initial $B$ meson and final  pseudoscalar $\pi$ or  vector $\rho$ mesons are usually parametrized by the following set of invariant form factors.

(a) $B \to \pi$ weak decays 
\begin{equation}
  \label{eq:pff1}
  \langle \pi(p_{\pi})|\bar q \gamma^\mu b|B(p_{B})\rangle
  =f_+(q^2)\left[p_{B}^\mu+ p_{\pi}^\mu-
\frac{M_{B}^2-M_{\pi}^2}{q^2}\ q^\mu\right]+
  f_0(q^2)\frac{M_{B}^2-M_{\pi}^2}{q^2}\ q^\mu,
\end{equation}
\begin{equation}
\label{eq:pff12} 
 \langle \pi(p_{\pi})|\bar q \gamma^\mu\gamma_5 b|B(p_{B})\rangle
 =\langle \pi(p_\pi)|\bar q \sigma^{\mu\nu}\gamma_5 q_\nu  b|B(p_{B})\rangle=0,
\end{equation}
\begin{equation}
\label{eq:pff2}
\langle \pi(p_\pi)|\bar q \sigma^{\mu\nu}q_\nu b|B(p_{B})\rangle=
\frac{if_T(q^2)}{M_{B}+M_\pi} [q^2(p_{B}^\mu+p_\pi^\mu)-(M_{B}^2-M_\pi^2)q^\mu],
\end{equation}

(b) $B \to \rho$ weak decays
\begin{eqnarray}
  \label{eq:vff1}
  \langle {\rho}(p_{\rho})|\bar q \gamma^\mu b|B(p_{B})\rangle&=
  &\frac{2iV(q^2)}{M_{B}+M_{\rho}} \epsilon^{\mu\nu\tau\sigma}\epsilon^*_\nu
  p_{B\tau} p_{{\rho}\sigma},\\ \cr
\label{eq:vff2}
\langle {\rho}(p_{\rho})|\bar q \gamma^\mu\gamma_5 b|B(p_{B})\rangle&=&2M_{\rho}
A_0(q^2)\frac{\epsilon^*\cdot q}{q^2}\ q^\mu
 +(M_{B}+M_{\rho})A_1(q^2)\left(\epsilon^{*\mu}-\frac{\epsilon^*\cdot
    q}{q^2}\ q^\mu\right)\cr\cr
&&-A_2(q^2)\frac{\epsilon^*\cdot q}{M_{B}+M_{\rho}}\left[p_{B}^\mu+
  p_{\rho}^\mu-\frac{M_{B}^2-M_{\rho}^2}{q^2}\ q^\mu\right], 
\end{eqnarray}
\begin{equation}
  \label{eq:vff3}
\langle  \rho(p_{\rho})|\bar q i\sigma^{\mu\nu}q_\nu b|B(p_{B})\rangle=2T_1(q^2)
\epsilon^{\mu\nu\tau\sigma} \epsilon^*_\nu p_{{\rho}\tau} p_{{B}\sigma},
\end{equation}
\begin{eqnarray}
\label{eq:vff4}
\langle \rho(p_{\rho})|\bar q i\sigma^{\mu\nu}\gamma_5q_\nu b|B(p_{B})\rangle&=&
T_2(q^2)[(M_{B}^2-M_{\rho}^2)\epsilon^{*\mu}-(\epsilon^*\cdot q)(p_{B}^\mu+
p_{\rho}^\mu)]\cr\cr
&&+T_3(q^2)(\epsilon^*\cdot q)\left[q^\mu-\frac{q^2}{M_{B}^2-M_{\rho}^2}
  (p_{B}^\mu+p_{\rho}^\mu)\right],
\end{eqnarray}
where $q=p_{B}- p_{\pi(\rho)}$ is the four-momentum transfer,
$M_{B,\pi(\rho)}$ are the initial and final meson masses, and  $\epsilon_\mu$ is the polarization vector of the final $\rho$ meson.

At the maximum recoil point ($q^2=0$) these form
factors satisfy the following conditions: 
\[f_+(0)=f_0(0),\]
\[A_0(0)=\frac{M_{B}+M_\rho}{2M_{\rho}}A_1(0)
-\frac{M_{B}-M_{\rho}}{2M_{\rho}}A_2(0),\]
\[T_1(0)=T_2(0).\]

In this paper we use the relativistic quark model based on the quasipotential approach and QCD for the calculation of the form factors of weak $B$ decays to final $\pi$ or $\rho$ mesons. The meson is described by the covariant single-time wave function which satisfy  the three-dimensional relativistically invariant Schr\"odinger-like
equation with the QCD-motivated interquark potential \cite{mass}  
\begin{equation}
\label{quas}
{\left(\frac{b^2(M)}{2\mu_{R}}-\frac{{\bf
p}^2}{2\mu_{R}}\right)\Psi_{M}({\bf p})} =\int\frac{d^3 q}{(2\pi)^3}
 V({\bf p,q};M)\Psi_{M}({\bf q}),
\end{equation}
where the relativistic reduced mass is
\begin{equation}
\mu_{R}=\frac{E_1E_2}{E_1+E_2}=\frac{M^4-(m^2_1-m^2_2)^2}{4M^3},
\end{equation}
with the on-mass-shell energies
\[E_1=\frac{M^2-m_2^2+m_1^2}{2M}, \quad E_2=\frac{M^2-m_1^2+m_2^2}{2M},
\]
 and $M=E_1+E_2$ is the meson mass, $m_{1,2}$ are the quark masses,
and ${\bf p}$ is their relative three-momentum.  
In the centre of mass system the on-mass-shell relative momentum squared 
$b^2(M)$ is expressed through the meson and quark masses:
\begin{equation}
{b^2(M) }
=\frac{[M^2-(m_1+m_2)^2][M^2-(m_1-m_2)^2]}{4M^2}.
\end{equation}

The kernel $V({\bf p,q};M)$ of  Eq.~(\ref{quas}) is the
quasipotential operator of the quark-antiquark interaction. It is constructed with the help of the off-mass-shell
scattering amplitude, projected onto the positive-energy states.
The explicit expression for the corresponding quasipotential
$V({\bf p,q};M)$  can be 
found in Ref.~\cite{mass}.

The constituent quark masses  $m_c=1.55$ GeV, $m_b=4.88$ GeV,
$m_u=m_d=0.33$ GeV, $m_s=0.5$ GeV and the parameters of the linear confining
potential $A=0.18$ GeV$^2$ and $B=-0.3$~GeV have been fixed previously
and have values typical for quark models. The value of the mixing coefficient of
vector and scalar confining potentials $\varepsilon=-1$ has been
determined from the consideration of charmonium radiative decays
\cite{mass} and  heavy quark effective theory. The universal
Pauli interaction constant $\kappa=-1$ has been fixed from the
analysis of the fine splitting of heavy quarkonia ${ }^3P_J$ -
states \cite{mass}. In this case, the long-range chromomagnetic
quark moment $(1+\kappa)$ vanishes in accordance with the flux-tube
model.

The matrix element of the weak current between the $B$ meson with mass $M_{B}$ and momentum $p_{B}$ and a final ($F=\pi$ or $\rho$) meson with mass $M_{F}$ and momentum $p_{F}$ is given  \cite{mass} by the expression
\begin{equation}\label{mxet} 
\langle F(p_{F}) \vert J^W_\mu \vert B(p_{B})\rangle
=\int \frac{d^3p\, d^3q}{(2\pi )^6} \bar \Psi_{{F}\,{\bf p}_F}({\bf
p})\Gamma _\mu ({\bf p},{\bf q})\Psi_{B\,{\bf p}_{B}}({\bf q}),
\end{equation}
where ${\bf p},{\bf q}$ are relative quark momenta, $\Gamma _\mu ({\bf p},{\bf
q})$ is the two-particle vertex function. Here
$\Psi_{M\,{\bf p}_M}({\bf p})$ are the meson ($M=B,{F})$ wave functions projected onto the positive energy states of
quarks ($q_1=b,u,d$ and $q_2=u,d$) and boosted to the moving reference frame with momentum ${\bf p}_M$
\begin{equation}
\label{wig}
\Psi_{M\,{\bf p}_M}({\bf
p})=D_{q_1}^{1/2}(R_{L_{{\bf p}_M}}^W)D_{q_2}^{1/2}(R_{L_{{
\bf p}_M}}^W)\Psi_{M\,{\bf 0}}({\bf p}),
\end{equation}
where $\Psi_{M\,{\bf 0}}({\bf p})\equiv \Psi_M({\bf p})$ is the wave function at rest, $R^W$ is the Wigner rotation, $L_{\bf\Delta}$ is the Lorentz boost
from the meson rest frame to a moving one and $D^{1/2}(R)$ is  
the spin rotation matrix.

Calculating the weak decay matrix elements we take into account both
the leading (spectator) term $\Gamma^{(1)}({\bf p},{\bf q})$ of
the vertex function  and subleading  term  $\Gamma^{(2)}({\bf
  p},{\bf q})$ which takes into account contributions of the
intermediate negative-energy states. The diagrams and explicit expressions for
these terms can be found in Refs.~\cite{rarebk,rarebs}. The previously developed methods
\cite{asld} allow us to express the relativistic decay matrix
elements through the usual overlap integrals of the initial and final
meson wave functions in their rest frames. These wave functions are
known from the meson mass spectrum calculations \cite{hlm,lm}. Note
that all considerations were done completely relativistically without
employing $v/c$ expansion. It is important to point out that the
obtained expressions for the decay matrix elements are valid in the
whole kinematical $q^2$ range accessible in weak decays. This fact allows one
to explicitly determine the $q^2$ dependence of  meson form factors
without additional model assumptions and extrapolations, thus
increasing reliability  of results.  The analytic expressions for
the form factors are given in Refs.~\cite{bdecay,rarebk}. It is
important to emphasize that these form factors in the heavy quark and
large recoil limits satisfy all model-independent relations imposed by heavy quark and
large energy effective theories \cite{orsay,efglr}.  

\begin{figure}
\centering
  \includegraphics[width=8cm]{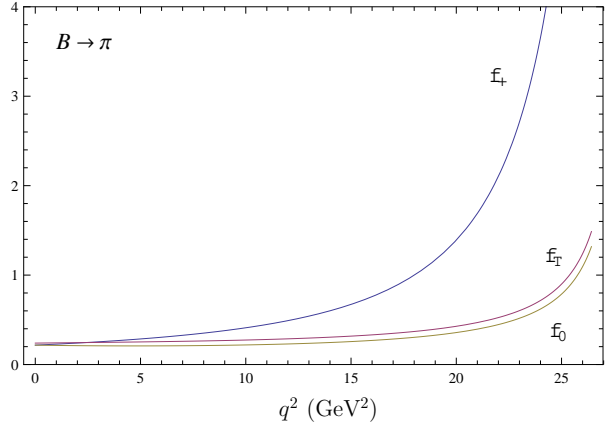}
\caption{Form factors of the weak $B\to \pi$ transition.    } 
\label{fig:fffpi}
\end{figure}

\begin{figure}
\centering
  \includegraphics[width=8cm]{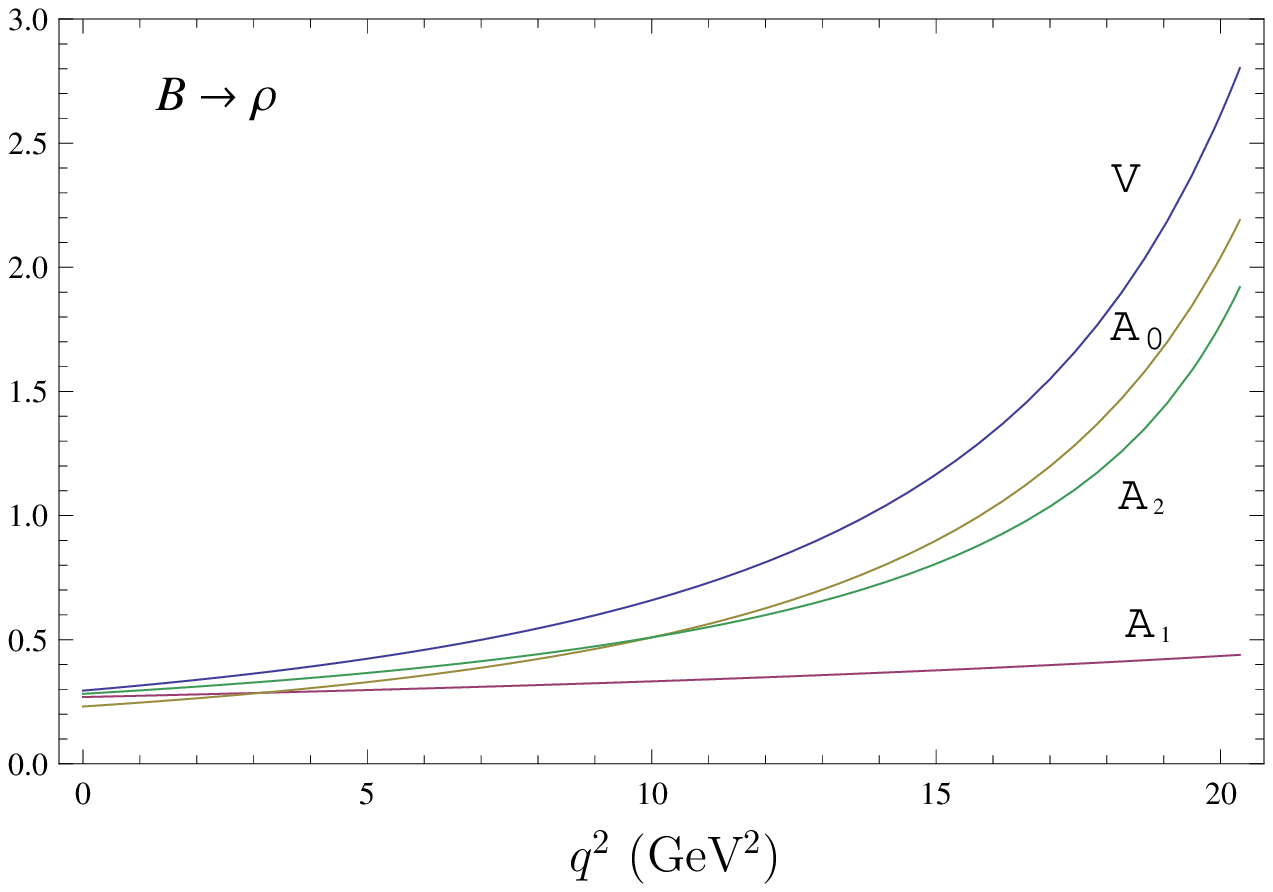}\ \
 \ \includegraphics[width=8cm]{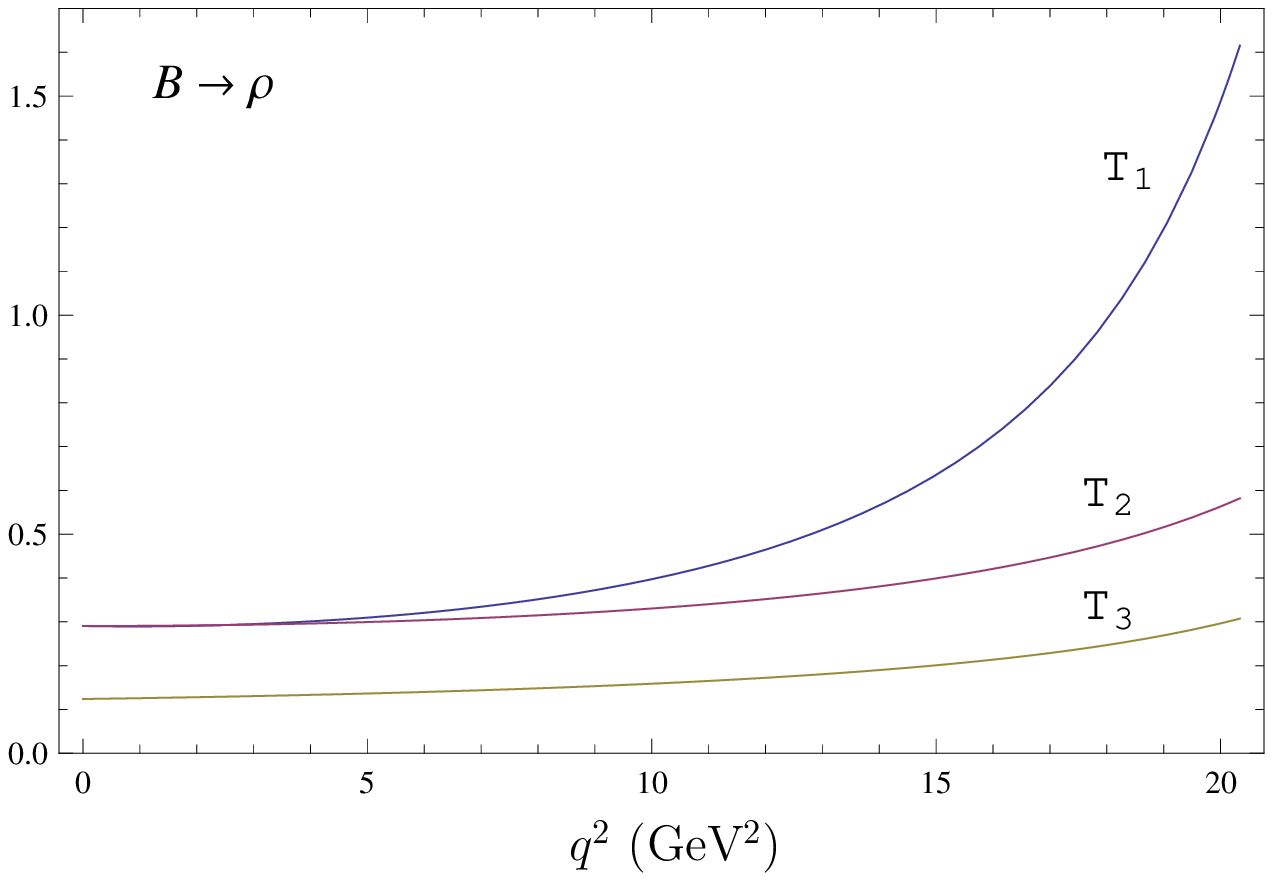}\\
\caption{Form factors of the weak $B\to \rho$ transition.    } 
\label{fig:fffrho}
\end{figure}

The numerical values of the form factors $f_+(q^2)$, $f_0(q^2)$, $V(q^2)$, $A_i(q^2)$ ($i=0,1,2$) parametrizing matrix elements of vector and axial vector weak currents for the $B\to\pi$ and $B\to\rho$ transitions were previously calculated in Ref.~\cite{asld}. Here we further extend our calculation to the form factors $f_T(q^2)$ and $T_i(q^2)$ ($i=1,2,3$) parametrizing matrix elements of the tensor and pseudotensor currents responsible for the rare $B\to\pi(\rho)l\bar l$ decays.  These form factors are plotted in Figs.~\ref{fig:fffpi} and \ref{fig:fffrho}.      

To simplify the comparison of the obtained form factors with experiment and
other theoretical calculations it is useful to have approximate analytic
expressions for them. Our analysis shows that the weak $B\to\pi(\rho)$ transition form factors can be well fitted  by the following formulas \cite{ms,asld}: 

(a) $F(q^2)= \{f_+(q^2),f_T(q^2),V(q^2),A_0(q^2),T_1(q^2)\}$ 
\begin{equation}
  \label{fitfv}
  F(q^2)=\frac{F(0)}{\displaystyle\left(1-\frac{q^2}{M^2}\right)
    \left(1-\sigma_1 
      \frac{q^2}{M_{B^*}^2}+ \sigma_2\frac{q^4}{M_{B^*}^4}\right)},
\end{equation}

(b) $F(q^2)=\{f_0(q^2), A_1(q^2),A_2(q^2),T_2(q^2),T_3(q^2)\}$
\begin{equation}
  \label{fita12}
  F(q^2)=\frac{F(0)}{\displaystyle \left(1-\sigma_1
      \frac{q^2}{M_{B^*}^2}+ \sigma_2\frac{q^4}{M_{B^*}^4}\right)},
\end{equation}
where $M=M_{B^*}$ for the form factors $f_+(q^2),f_T(q^2),V(q^2),T_1(q^2)$ and
$M=M_{B}$ for the form factor $A_0(q^2)$. The obtained values of $F(0)$ and
$\sigma_{1,2}$ are given in Table~\ref{hff}. The accuracy of such approximation is
rather high, the deviation from the
calculated form factors does not exceed 1\%. The rough
estimate of the total uncertainty of the form factors within
our model is of order of 5\%.  

\begin{table}
\caption{The form factors of weak $B\to \pi(\rho)$ weak transitions.
}
\label{hff}
\begin{ruledtabular}
\begin{tabular}{ccccccccccc}
   &\multicolumn{3}{c}{{$B\to \pi$}}&\multicolumn{7}{c}{{\  $B\to \rho$\
     }}\\
\cline{2-4} \cline{5-11}
& $f_+$ & $f_0$& $f_T$& $V$ & $A_0$ &$A_1$&$A_2$& $T_1$ & $T_2$& $T_3$\\
\hline
$F(0)$          &0.217 &0.217 &  0.240 & 0.295 & 0.231& 0.269 & 0.282 & 0.290& 0.290& 0.124\\
$F(q^2_{\rm max})$&10.9  &1.32 &  1.64 & 2.80& 2.19& 0.439 &  1.92 & 1.62& 0.582& 0.307\\
$\sigma_1$      &$0.378$&$-0.501$& $-1.19$& $0.875$ &$0.796$&  0.540&
1.34  &$-1.21$&$0$& $0.423$\\
$\sigma_2$      &$-0.410$&$-1.50$&$0.047$&$0$&$-0.055$&$0$&
0.210  &$-2.40$&$-0.974$& $-0.571$\\
\end{tabular}
\end{ruledtabular}
\end{table}

\begin{sidewaystable}
\caption{Comparison of theoretical predictions for the form factors of 
  weak $B\to  \pi(\rho)$  transitions at the maximum
  recoil point $q^2=0$.  }
\label{compbff}
\begin{ruledtabular}
\begin{tabular}{ccccccccc}
     & $f_+(0)$& $f_T(0)$ & $V(0)$ & $A_0(0)$ &$A_1(0)$&$A_2(0)$ &$T_1(0)$&$T_3(0)$ \\
\hline
This paper   &$0.217\pm0.011$  & $0.240\pm0.012$  &$0.295\pm0.015$  &$0.231\pm0.012$ &$0.269\pm0.014$  &$0.282\pm0.014$ & $0.290\pm0.015$& $0.124\pm0.007$\\
\cite{bz}   & $0.258\pm0.031$ &$0.253\pm0.028$&$0.323\pm0.030$ & $0.303\pm0.029$ & $0.242\pm0.029$& $0.221\pm0.023$& $0.267\pm0.023$& $0.176\pm0.016$\\
\cite{kmo} &$0.25\pm0.05$  &$0.21\pm0.04$&$0.32\pm0.10$& &$0.24\pm0.08$&
$0.21\pm0.09$&$0.28\pm0.09$\\
\cite{ms}& 0.29&0.28 &$0.31$ &0.30 & 0.26 &0.24&0.27&0.19 \\
 \cite{ikkss} &0.29 & 0.27&0.28 & & 0.26 & 0.24 & 0.25\\
\cite{lww} &0.247 & 0.253 &0.298&0.260 &0.227&0.215&0.260&0.184\\
\cite{wx} & $0.26\pm0.05$& $0.26\pm0.05$\\
\cite{apr} &$0.261\pm0.014$ &$0.231\pm0.013$ 
\end{tabular}
\end{ruledtabular}
\end{sidewaystable}

In Table~\ref{compbff} we compare the predictions of our model for the
weak $B\to  \pi(\rho)$ decay form factors at the maximum recoil point
$q^2=0$ with other theoretical calculations
\cite{bz,kmo,ms,ikkss,lww,wx,apr}. The different versions of
light-cone sum rules are employed in Refs.~\cite{bz,kmo}. Calculations
of Ref.~\cite{ms} are based on the constituent quark model within
relativistic dispersion approach while in Ref.~\cite{ikkss} 
the covariant constituent quark model with the infrared confinement is
applied. The perturbative QCD factorization approach with the
inclusion of the leading and next-to-leading-order corrections is used
in Refs.~\cite{lww,wx}. The authors of Ref.~\cite{apr} extract the
form factors combining available experimental data on
semileptonic $B\to \pi l\nu_l$ decays and lattice QCD calculations of
the corresponding form factors for the rare $B\to K$ transitions
within the $SU(3)$-breaking Ansatz.  Comparison of the results presented
in this table shows that, although there are some 
differences between the central values of predictions, in general there is a reasonable
agreement between the values of these form factors at zero recoil
calculated using significantly different theoretical methods.  

Note that most of the discussed theoretical approaches allow to calculate the form factors at a single point only or in some limited range of the recoil momentum, then some model extrapolation to the whole kinematical range should be used.
The important advantage of our approach consists in the fact, that it determines various decay form factors through the overlap
integrals of hadron wave functions  in the whole kinematically accessible
range without additional assumptions and extrapolations. These wave
functions are obtained by numerical solving  equation (1) with the
nonperturbative treatment of relativistic effects. 

We can further test our model  confronting the form factor $f _0$ at zero recoil point ($q^2=q^2_{\rm max}$) with the Callan-Treiman-type normalization condition 
$$f_0(q^2_{\rm max})\approx \frac{f_B}{f_\pi}$$ 
derived using  the soft pion  limit $p\to 0$ and $M_\pi^2\to 0$ \cite{dks}. Taking our prediction for the decay constant $f_B=189$~MeV \cite{dconst}, which is well consistent with the averaged theoretical value given in Ref.~\cite{pdg}, and the experimental value of $f_\pi$ \cite{pdg} we get  $f_0(q^2_{\rm max})\approx1.45$ in good agreement with the value 1.32 given in Table~\ref{hff} (see also Ref.~\cite{asld}).

\section{Semileptonic $B\to \pi l\nu_l$ and $B\to \rho l\nu_l$ decays}
\label{sec:sld}

We start from the consideration of the semileptonic $B\to \pi l\nu_l$ and $B\to \rho l\nu_l$ decays. They were investigated in detail in Ref.~\cite{asld}, where all necessary formulas and values of branching fractions can be found.   Using the new data from Belle \cite{Belle1,Belle2} and BaBar
\cite{Babar} on the exclusive charmless semileptonic $B$ decays we can
update our analysis in Ref.~\cite{asld} as follows.

The branching fractions of such decays predicted by our model are given
\cite{asld} by 
  \begin{eqnarray}
  \label{eq:brtot1}
  Br(\bar B^0\to \pi^+ l^-\nu)&=&8.36|V_{ub}|^2,\cr
Br(B^-\to \pi^0 l^-\nu)&=&4.51|V_{ub}|^2,\cr
 Br(\bar B^0\to \rho^+ l^-\nu)&=&20.12|V_{ub}|^2,\cr
Br(B^-\to \rho^0 l^-\nu)&=&10.87|V_{ub}|^2.
\end{eqnarray}  


Comparing these predictions with recent experimental data
\cite{Belle1,Belle2,Babar} 
 \begin{eqnarray}
  \label{eq:brexp}
  Br(\bar B^0\to \pi^+ l^-\nu)&=&(1.49\pm0.04\pm0.07)\times
  10^{-4}\ [1],\cr Br(\bar B^0\to \pi^+
  l^-\nu)&=&(1.49\pm0.09\pm0.07)\times 10^{-4}\ [2],\cr
  Br(\bar B^0\to \pi^+ l^-\nu)&=&(1.47\pm0.05\pm0.06)\times 10^{-4}\ [3],\cr
Br(B^-\to \pi^0 l^-\nu)&=&(0.80\pm0.08\pm0.04)\times 10^{-4}\
[2],\cr
Br(B^-\to \pi^0 l^-\nu)&=&(0.77\pm0.04\pm0.03)\times 10^{-4}\ [3],\cr
 Br(B^0\to \rho^+ l^-\nu)&=&(3.22\pm0.27\pm0.24)\times 10^{-4}\ [2],\cr
Br(B^-\to \rho^0 l^-\nu)&=&(1.83\pm0.10\pm0.10)\times 10^{-4}\ [2].
\end{eqnarray} 
we find the values of the CKM matrix element $|V_{ub}|$ presented
 in Table~\ref{vub}.

\begin{table}
\caption{Values of the CKM matrix element $|V_{ub}|\times 10^3$ extracted in our
  model from the recent experimental data. Only experimental errors
  are given.}
\label{vub}
\begin{ruledtabular}
\begin{tabular}{cccc}
 Decay& Belle untagged \cite{Belle1}& Belle tagged  \cite{Belle2}  &BaBar \cite{Babar}\\
\hline
$\bar B^0\to\pi^+l^-\nu_l$& $4.22\pm0.12$& $4.22\pm0.14$& $4.19\pm0.11$\\
$B^-\to\pi^0l^-\nu_l$&& $4.21\pm0.23$&  $4.14\pm0.14$\\
$\bar B^0\to\rho^+l^-\nu_l$&& $4.00\pm0.24$ & \\
$B^-\to\rho^0l^-\nu_l$&& $4.10\pm0.16$ & \\
\end{tabular}
\end{ruledtabular}
\end{table}

Averaging these values we get the following exclusive value for the CKM matrix element  $|V_{ub}|$
\begin{equation}
  \label{eq:vub}
  |V_{ub}|=(4.15\pm0.09_{\rm exp}\pm0.21_{\rm theor})\times 10^{-3} \qquad ({\rm exclusive}) ,
\end{equation}
where the last error is the rough (conservative upper) estimate of the
theoretical uncertainties within  our model. Note that this value is
consistent with the one extracted from the inclusive charmless
semileptonic $B$ decays \cite{pdg}
\begin{equation}
  \label{eq:vubinc}
  |V_{ub}|=(4.41\pm0.15^{+0.15}_{-0.17})\times 10^{-3} \qquad ({\rm inclusive}) .
\end{equation}

\begin{figure}
\centering
  \includegraphics[width=10cm]{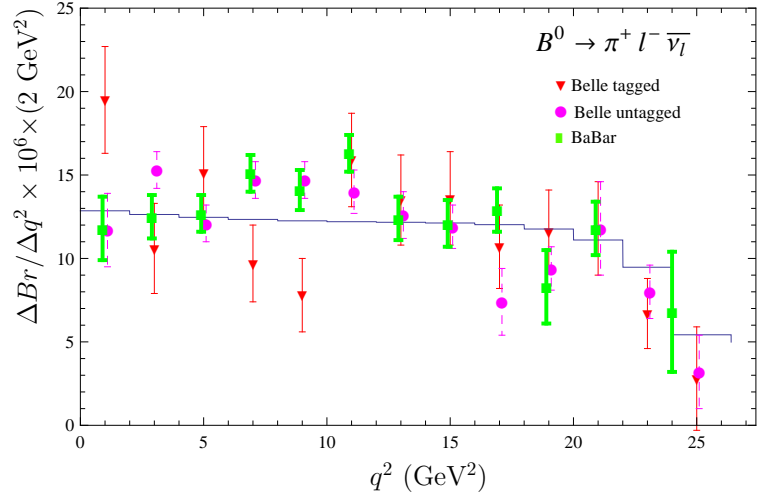}
\caption{Comparison of predictions  of our model with  recent
  experimental data for the $B^0\to \pi^+ l^-\nu$ decay \cite{Belle1,Belle2,Babar}.    } 
\label{fig:ffppl}
\end{figure}

\begin{figure}
\centering
  \includegraphics[width=8cm]{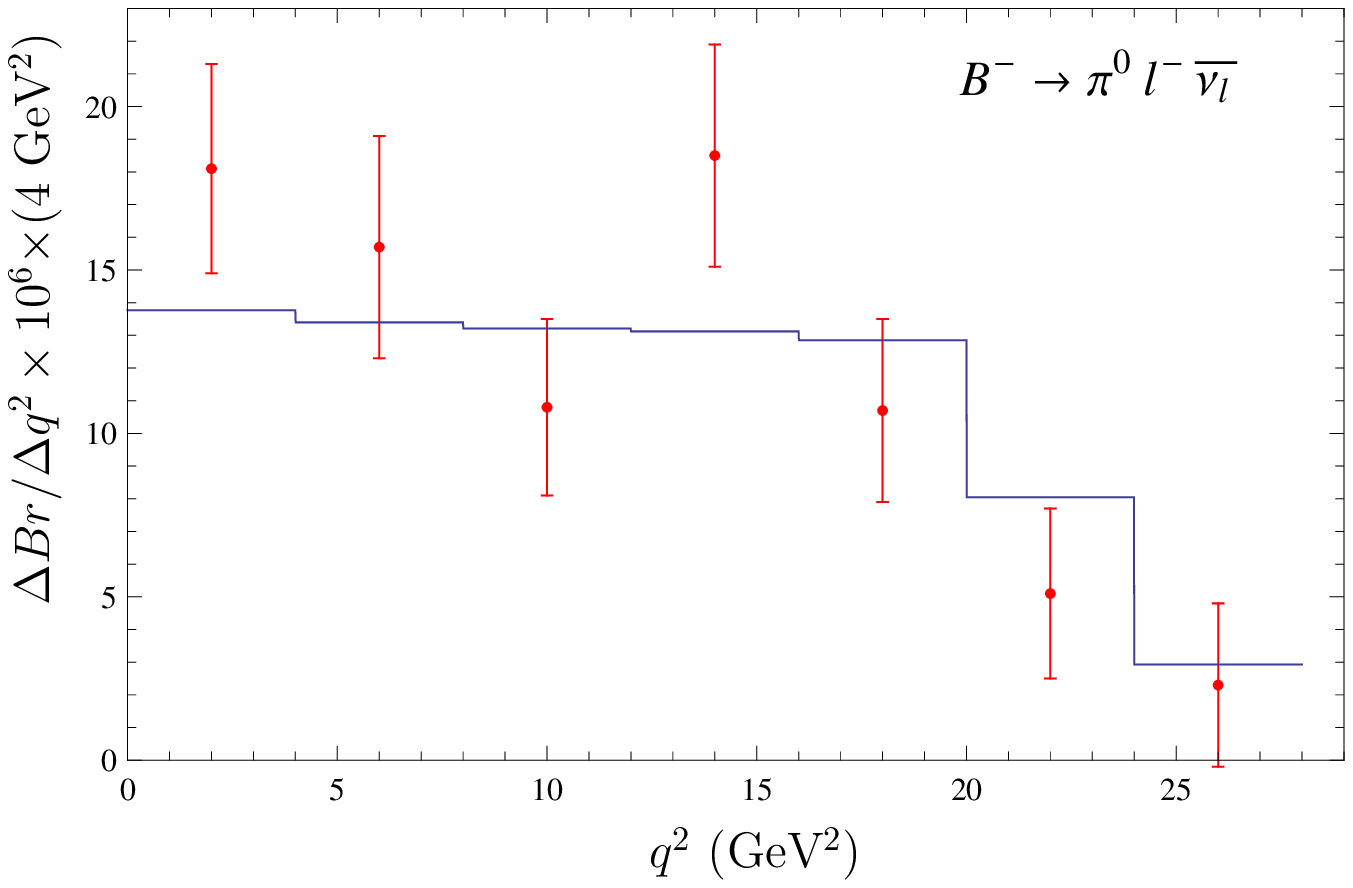}\ \
 \ \includegraphics[width=8cm]{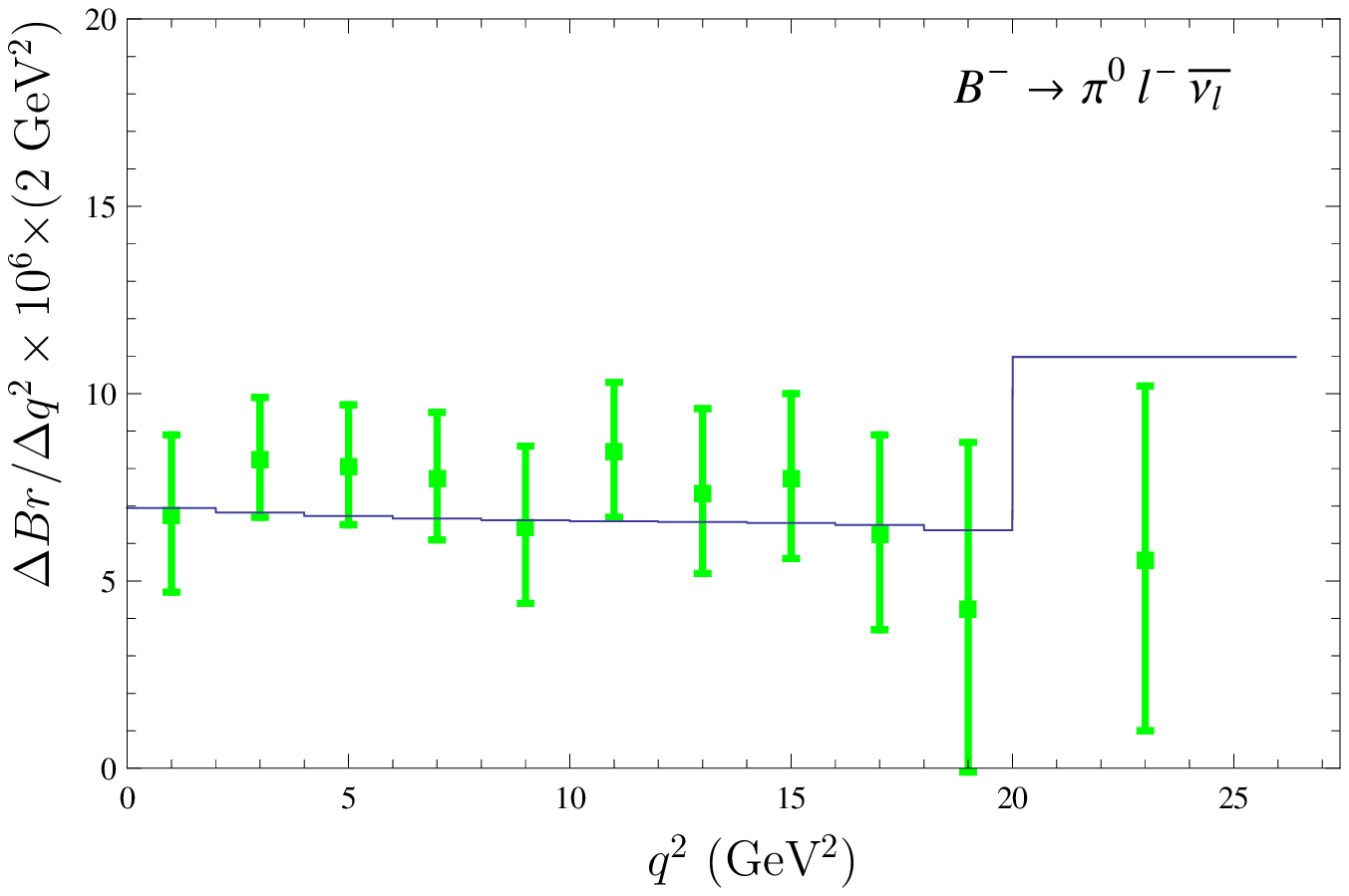}\\
\caption{Comparison of predictions  of our model with  recent
  experimental data for the $B^-\to \pi^0 l^-\nu$ decay (left figure --
  Belle data \cite{Belle1}, right figure -- BaBar data \cite{Babar}).    } 
\label{fig:ffp0}
\end{figure}

\begin{figure}
\centering
  \includegraphics[width=8cm]{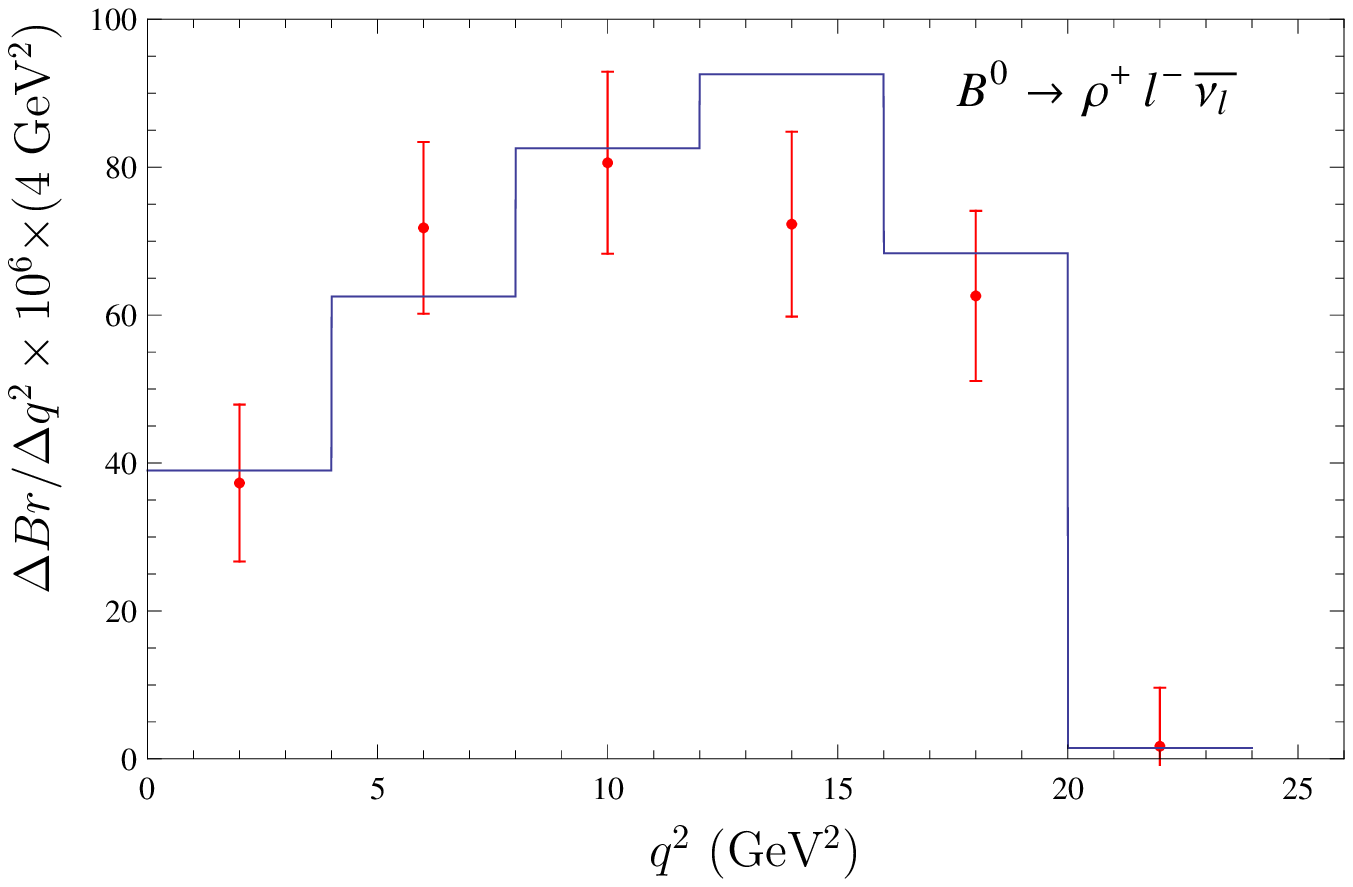}\ \
 \ \includegraphics[width=8cm]{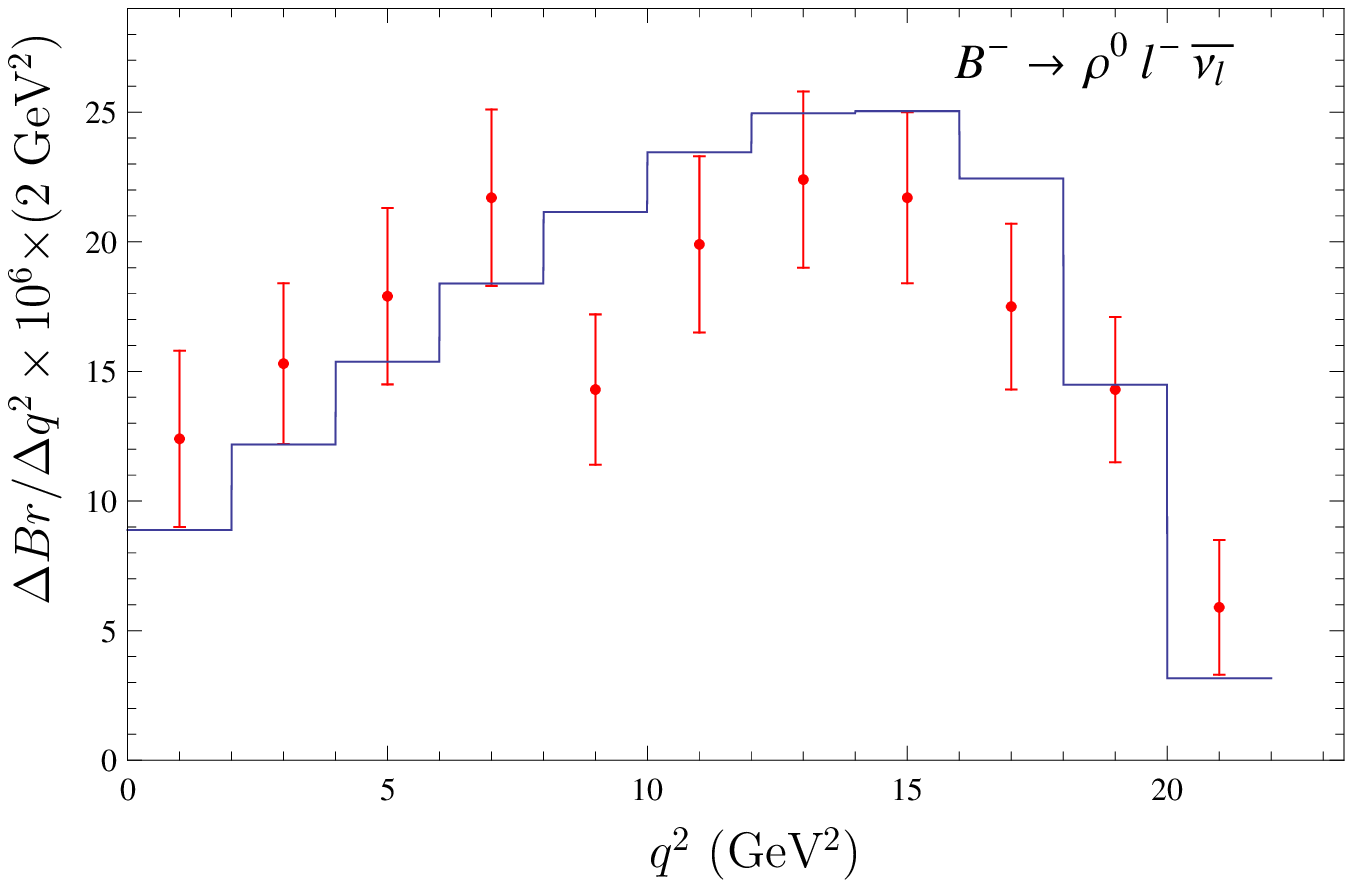}\\
\caption{Comparison of predictions  of our model with recent
  Belle data \cite{Belle1} for the $B\to \rho l\nu$ decay.    } 
\label{fig:ffrho}
\end{figure}

It is important to point out that recent data provide us not only
the total branching fractions but also the partial branching
fractions $\Delta Br/\Delta q^2$  averaged over rather small $q^2$ bins. This allows one to test rather
precisely the $q^2$ dependence of decay form factors. The comparison
of our results for the partial decay rates with recent data is given in
Figs.~\ref{fig:ffppl}--\ref{fig:ffrho}.  In Fig.~\ref{fig:ffppl} we
confront our predictions for the semileptonic decay of the neutral $\bar
B^0$ meson to the charged $\pi^+$ meson with recent untagged and tagged
data from Belle \cite{Belle1,Belle2} and data from BaBar \cite{Babar},
while in Fig.~\ref{fig:ffp0} the corresponding predictions and the Belle data
for the decay of the charged $B^-$ meson to the neutral $\pi^0$ meson
are presented. Differential branching fractions for decays of the charged and neutral $B$ mesons to $\rho$ mesons are plotted in Fig.~\ref{fig:ffrho}. Here only Belle data are available \cite{Belle2}. From these figures we see that the reasonable agreement of
our theoretical results and data is observed both for the
semileptonic $B$ decays to the pseudoscalar $\pi$ and vector $\rho$ mesons. 
In most cases our predictions agree with data within error bars or lie
just in-between individual measurements. This comparison assures  the
reliability of our approach utilizing the model form factors.

\section{Rare semileptonic $B\to \pi(\rho) l^+l^-$ and $B\to \pi(\rho)
  \nu\bar \nu$ decays }
\label{sec:rd}
Now we apply the calculated weak decay form factors to the
consideration of the rare $B$ decays to light $\pi$ or $\rho$
mesons. Such decays are significantly less studied
experimentally. Their theoretical description is
usually based on the effective Hamiltonian ${\cal H}_{\rm eff}$ in
which heavy degrees of freedom (gauge bosons and top quark) are
integrated out. The operator product
expansion allows the separation of short- and long-distance effects
which are assumed to factorize. The short-distance 
contributions are described by the Wilson coefficients $c_i$ which are
calculated within perturbation theory, while the long-distance part is
attributed to the set of the standard model operators ${\cal O}_i$.

The effective Hamiltonian for the $b\to d l^+ l^-$
transitions can be presented \cite{bhi}  in the following form  taking into account the unitarity of the CKM matrix    
\begin{equation}
  \label{eq:heff}
  {\cal H}_{\rm eff} =-\frac{4G_F}{\sqrt{2}}\left[V_{td}^*V_{tb}\sum_{i=1}^{10}c_i{\cal O}_i+ V_{ud}^*V_{ub}\sum_{i=1}^{2}c_i\left({\cal O}_i-{\cal O}_i^{(u)}\right) \right],
\end{equation}
where $G_F$ is the Fermi constant, $V_{tj}$ and $V_{uj}$ are the
CKM matrix elements, $c_i$ are the Wilson coefficients
and ${\cal O}_i({\cal O}_i^{(u)})$ comprise the four-quark operator basis. Then the resulting transition amplitude is given by
  \begin{eqnarray}
  \label{eq:mtl}
  {\cal M}&=&\frac{G_F\alpha}{\sqrt{2}\pi}
  |V_{td}^*V_{tb}|\Biggl\{(\bar d\left[c_9^{\rm eff}\gamma_\mu(1-\gamma_5)-
    \frac{2m_b}{q^2}c_7^{\rm eff}i\sigma_{\mu\nu}q^\nu(1+\gamma_5)\right]b)(\bar
    l\gamma^\mu l)\cr
&&+c_{10}(\bar d\gamma_\mu(1-\gamma_5)b)
    (\bar l\gamma^\mu \gamma_5l)\Biggr\}.
\end{eqnarray}
The values of the Wilson coefficients $c_i$ and of the effective
Wilson coefficient $c_7^{\rm eff}$  are taken from Ref.~\cite{wc}. The 
effective  Wilson coefficient $ c_9^{\rm eff}$ contains additional
perturbative and long-distance contributions. It can be written as
\begin{eqnarray}
  \label{eq:c9eff}
 c_9^{\rm eff}&=&c_9+h^{\rm eff}\left(\frac{m_c}{m_b},\frac{q^2}{m_b^2}\right)(3c_1+c_2+3c_3+c_4+3c_5+c_6)\cr
&&+\lambda_u\left[h^{\rm eff}\left(\frac{m_c}{m_b},\frac{q^2}{m_b^2}\right)-h^{\rm eff}\left(\frac{m_u}{m_b},\frac{q^2}{m_b^2}\right)\right](3c_1+c_2)-
\frac12 h\left(1,\frac{q^2}{m_b^2}\right)(4c_3+4c_4+3c_5+c_6)\cr
&&-\frac12
h\left(0,\frac{q^2}{m_b^2}\right)(c_3+3c_4)+\frac29(3c_3+c_4+3c_5+c_6),
\end{eqnarray}   
where $\lambda_u=\frac{V_{ud}^*V_{ub}}{V_{td}^*V_{tb}}$ and
\begin{eqnarray*} 
h\left(\frac{m_c}{m_b},  \frac{q^2}{m_b}\right) & = & 
- \frac{8}{9}\ln\frac{m_c}{m_b} +
\frac{8}{27} + \frac{4}{9} x 
-  \frac{2}{9} (2+x) |1-x|^{1/2} \left\{
\begin{array}{ll}
 \ln\left| \frac{\sqrt{1-x} + 1}{\sqrt{1-x} - 1}\right| - i\pi, &
 x \equiv \frac{4 m_c^2}{ q^2} < 1,  \\
 & \\
2 \arctan \frac{1}{\sqrt{x-1}}, & x \equiv \frac
{4 m_c^2}{ q^2} > 1,
\end{array}
\right. \\
h\left(0, \frac{q^2}{m_b} \right) & = & \frac{8}{27} - 
\frac{4}{9} \ln\frac{q^2}{m_b} + \frac{4}{9} i\pi,
\end{eqnarray*}
while the function
\begin{equation}
  \label{eq:ybw}
h^{\rm
  eff}\left(\frac{m_c}{m_b},\frac{q^2}{m_b^2}\right)=h\left(\frac{m_c}{m_b},\frac{q^2}{m_b^2}\right)+
\frac{3\pi}{\alpha^2 c_0} \sum_{V_i=J/\psi,\psi(2S)\dots}\frac{\Gamma(V_i\to l^+l^-)M_{V_i} }{M_{V_i}^2-q^2-iM_{V_i}\Gamma_{V_i}}
\end{equation}
contains additional long-distance (nonperturbative) contributions which originate from the $c\bar c$ mesons [$J/\psi,
\psi(2S)\dots$]. We include contributions of the vector $V_i(1^{--})$ charmonium states: $J/\psi$,
$\psi(2S)$, $\psi(3770)$, $\psi(4040)$, $\psi(4160)$ and $\psi(4415)$,
with their masses ($M_{V_i}$), leptonic [$\Gamma(V_i\to l^+l^-)$] and
total ($\Gamma_{V_i}$) decay widths taken from PDG \cite{pdg}. The
coefficient $c_0=3c_1+c_2+3c_3+c_4+3c_5+c_6$. Similar expression holds
for the function $h^{\rm
  eff}\left(\frac{m_u}{m_b},\frac{q^2}{m_b^2}\right)$, where the
long-distance contributions now come from  $u\bar u$ states ($\rho$ and
$\omega$). 

The matrix element of the $b\to d l^+l^-$ transition amplitude between
meson states can be expressed through the helicity amplitudes
$H^{(i)}_m$ (where the superscript $i=1,2$ corresponds to the first and
second terms in the amplitude (\ref{eq:mtl}), while the subscript
$m=\pm,0,t$ denotes transverse, longitudinal and time helicity
components, respectively). The explicit  formulas for the helicity amplitudes in terms of the decay form factors defined in Eqs.~(\ref{eq:pff1})-(\ref{eq:vff4}) are given in our papers~\cite{rarebk,rarebs}.  

Then the differential decay rate can be written in terms of the helicity
amplitudes \cite{fgikl,rarebk} as follows.

(a) for the $B\to \pi(\rho)l^+l^-$ decays
\begin{eqnarray}
  \label{eq:dgamma}
  \frac{d\Gamma(B\to \pi(\rho)l^+l^-)}{dq^2}&=&\frac{G_F^2}{(2\pi)^3}
  \left(\frac{\alpha |V_{td}^*V_{tb}|}{2\pi}\right)^2
  \frac{\lambda^{1/2}q^2}{48M_{B}^3} \sqrt{1-\frac{4m_l^2}{q^2}}
  \Biggl[H^{(1)}H^{\dag(1)}\left(1+\frac{2m_l^2}{q^2}\right)\cr
&& +
    H^{(2)}H^{\dag(2)}\left(1-\frac{4m_l^2}{q^2}\right) +\frac{2m_l^2}{q^2}3 H^{(2)}_tH^{\dag(2)}_t\Biggr],
\end{eqnarray}

(b) for the $B\to \pi(\rho)\nu\bar\nu$ decays
\begin{equation}
  \label{eq:dgnunu}
  \frac{d\Gamma(B\to \pi(\rho)\nu\bar\nu)}{dq^2}=3\frac{G_F^2}{(2\pi)^3}
  \left(\frac{\alpha |V_{td}^*V_{tb}|}{2\pi}\right)^2
  \frac{\lambda^{1/2}q^2}{24M_{B}^3}
  H^{(\nu)}H^{\dag(\nu)},
\end{equation}
where $m_l$ is the lepton mass and
\begin{equation}
  \label{eq:hh}
  H^{(i)}H^{\dag(i)}\equiv H^{(i)}_+H^{\dag(i)}_++H^{(i)}_-H^{\dag(i)}_-+H^{(i)}_0H^{\dag(i)}_0.
\end{equation}

The forward-backward asymmetry for the $B\to\rho \mu^+\mu^-$ decay can
be expressed in terms of the helicity amplitudes in the following way
\begin{equation}
  \label{eq:afb}
  A_{FB}=\frac34 \sqrt{1-\frac{4m_l^2}{q^2}}\frac{ {\rm Re}(H^{(1)}_+H^{\dag(2)}_+)-{\rm Re}(H^{(1)}_-H^{\dag(2)}_-)}{H^{(1)}H^{\dag(1)}\left(1+\frac{2m_l^2}{q^2}\right) +
    H^{(2)}H^{\dag(2)}\left(1-\frac{4m_l^2}{q^2}\right) +\frac{2m_l^2}{q^2}3 H^{(2)}_tH^{\dag(2)}_t},
\end{equation}
while the longitudinal polarization fraction of the vector $\rho$
meson is given by
\begin{equation}
  \label{eq:fl}
  F_L=\frac{H^{(1)}_0H^{\dag(1)}_0\left(1+\frac{2m_l^2}{q^2}\right) +
    H^{(2)}_0H^{\dag(2)}_0\left(1-\frac{4m_l^2}{q^2}\right) +\frac{2m_l^2}{q^2}3 H^{(2)}_tH^{\dag(2)}_t }{H^{(1)}H^{\dag(1)}\left(1+\frac{2m_l^2}{q^2}\right) +
    H^{(2)}H^{\dag(2)}\left(1-\frac{4m_l^2}{q^2}\right) +\frac{2m_l^2}{q^2}3 H^{(2)}_tH^{\dag(2)}_t}.
\end{equation}

\begin{figure}
  \centering
   \includegraphics[width=7.8cm]{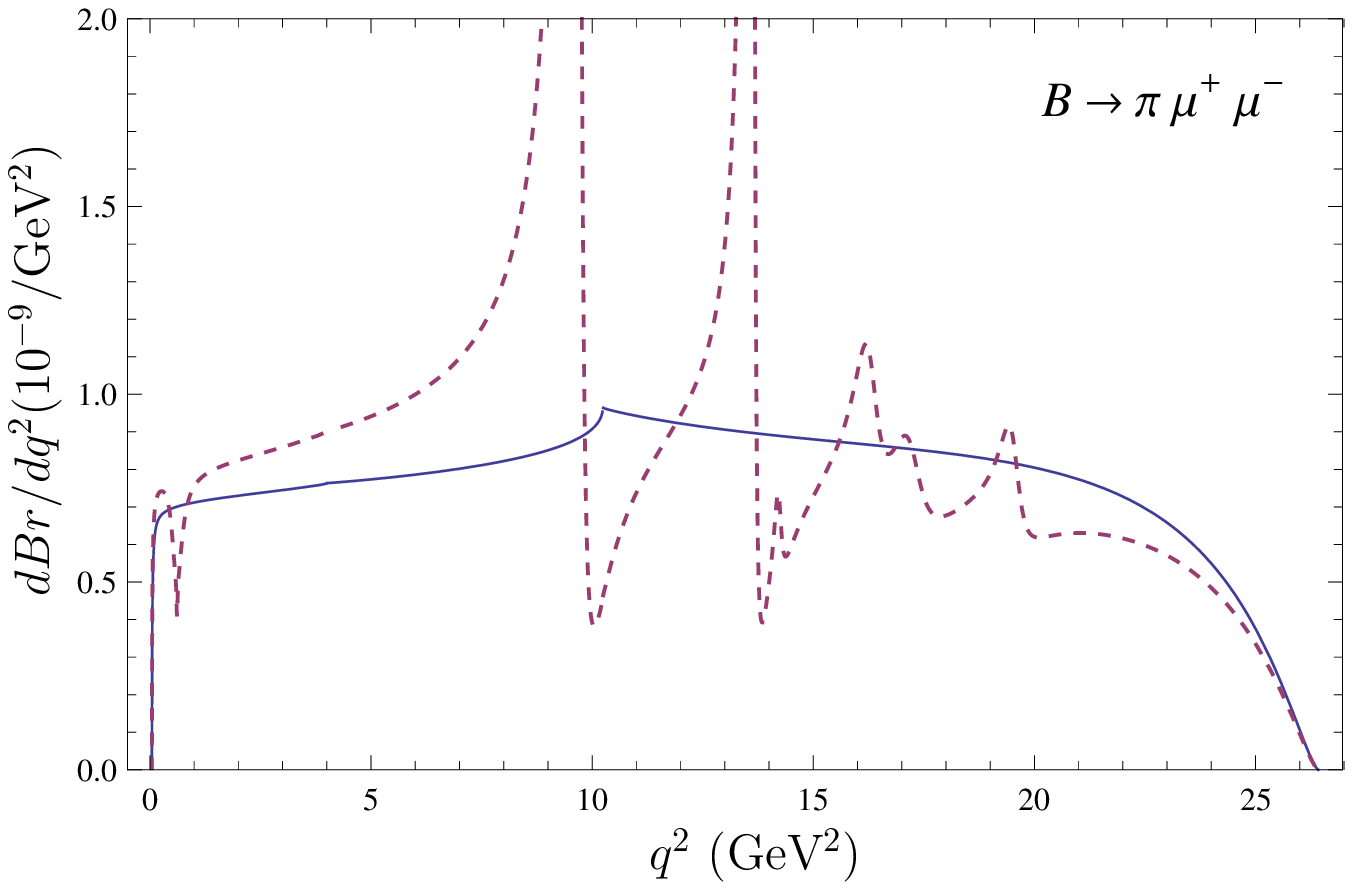}\ \ \
  \includegraphics[width=8cm]{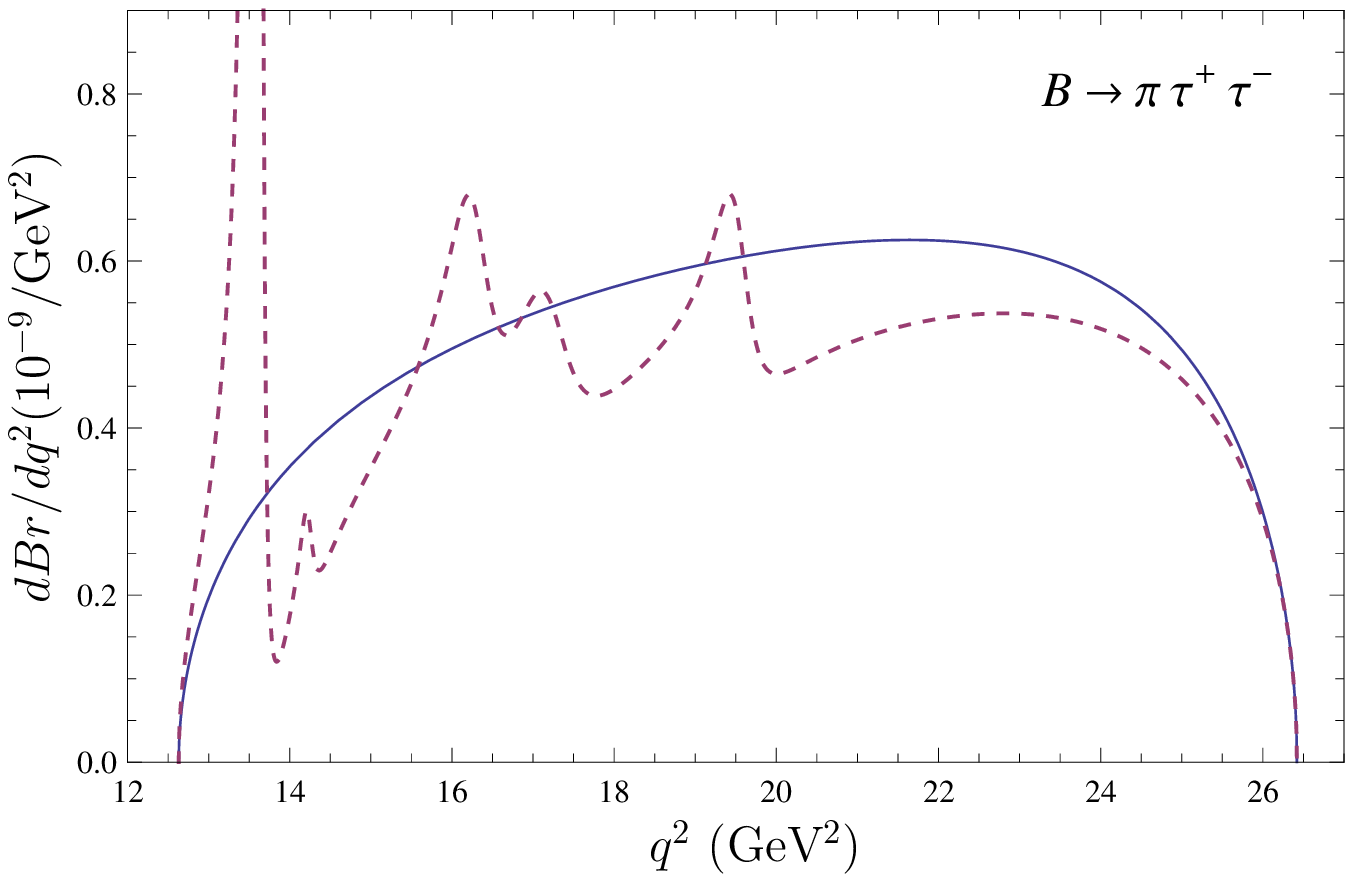}
  \caption{Theoretical predictions for the differential
    branching fractions $d Br(B^+ \to \pi^+ l^+l^-)/d q^2$. Nonresonant
    and resonant results are plotted by solid and dashed lines,
    respectively. }
  \label{fig:brbpi}
\end{figure}

\begin{figure}
  \centering
 \includegraphics[width=7.8cm]{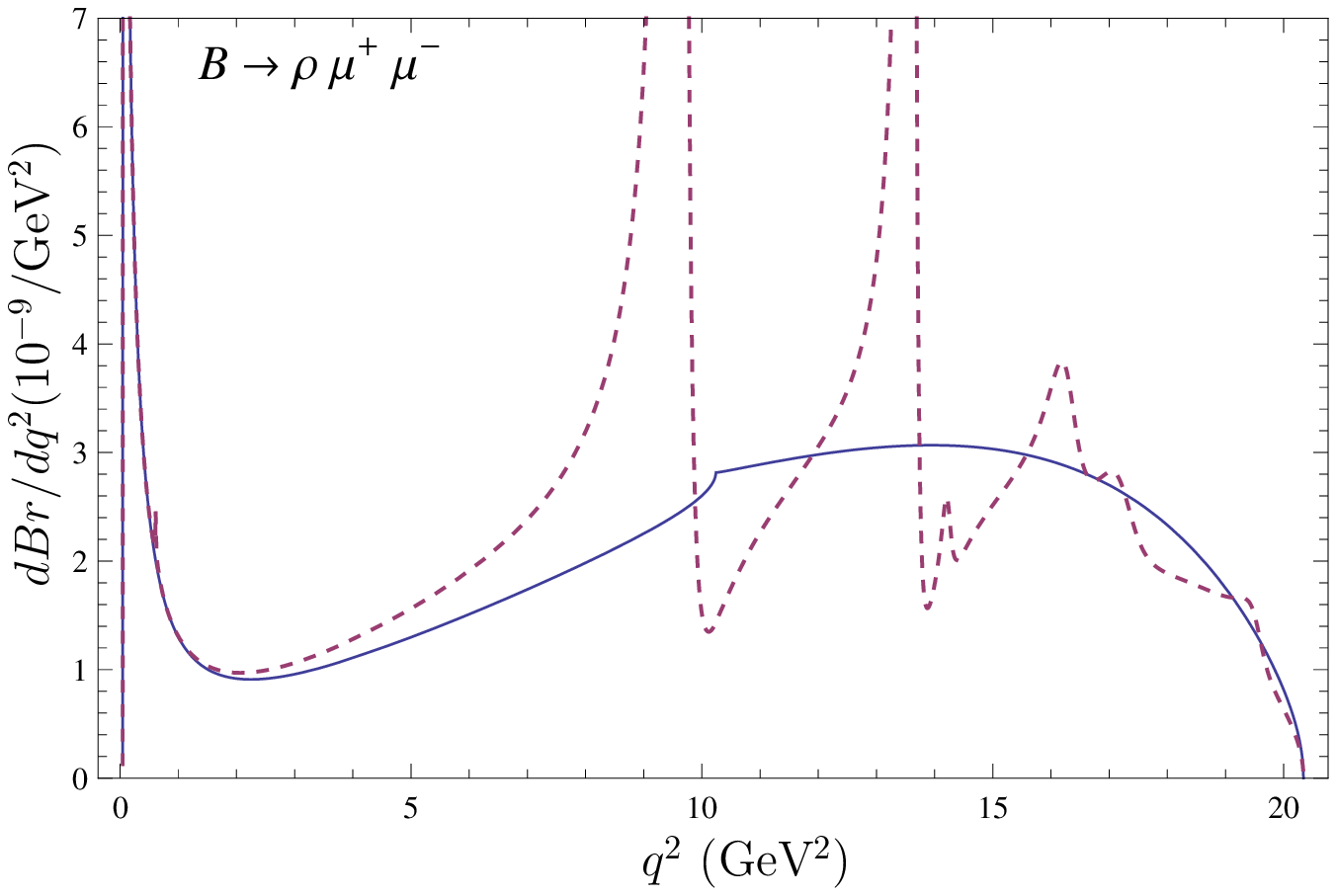} \ \ \  \includegraphics[width=7.8cm]{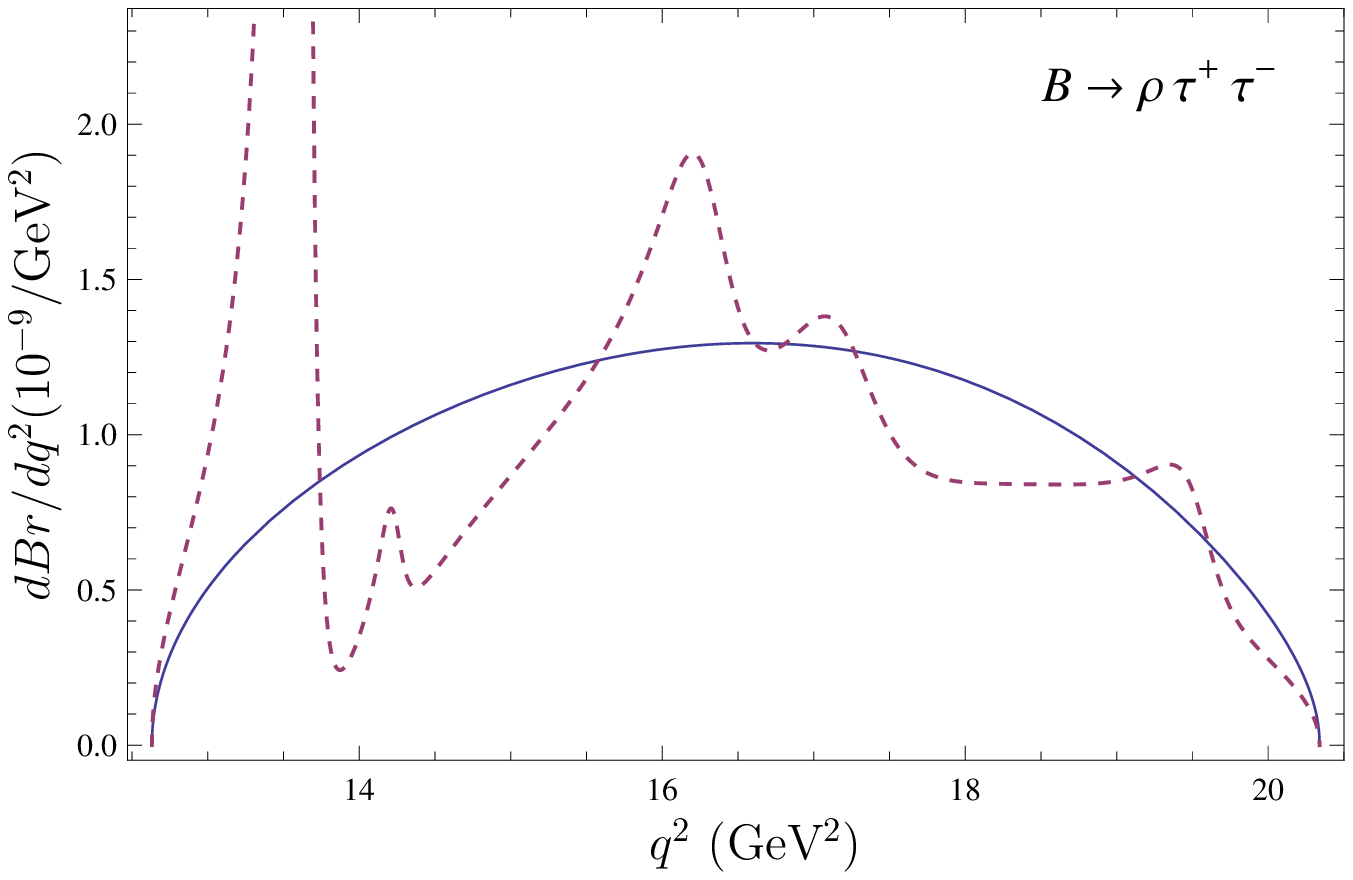}
\caption{Same as in Fig.~\ref{fig:brbpi} but for $d Br(B \to \rho
  l^+l^-)/d q^2$ decay. }
  \label{fig:brbrho}
\end{figure}

\begin{figure}
  \centering
\includegraphics[width=7.8cm]{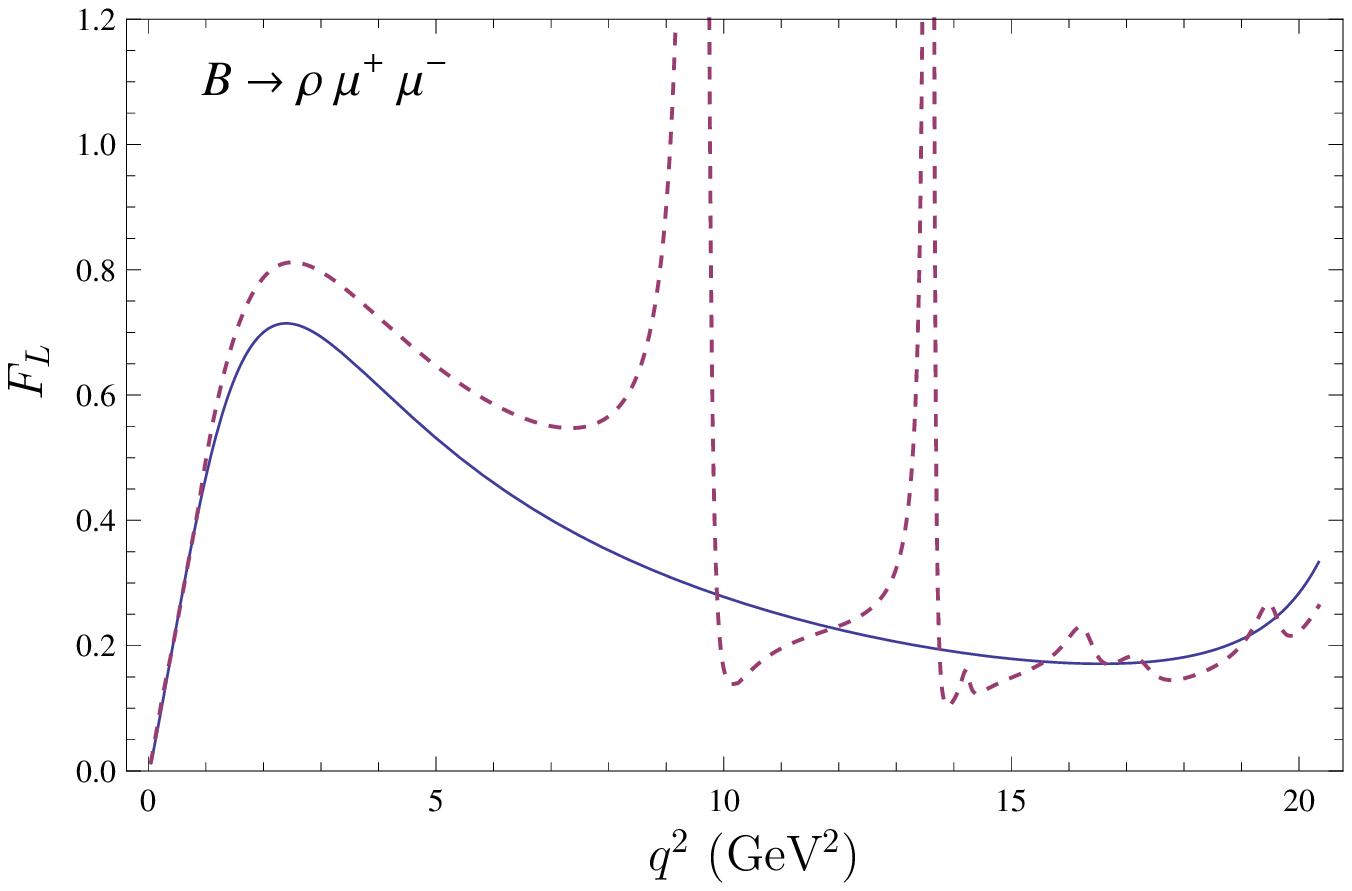}\ \ \ \
\  \includegraphics[width=7.8cm]{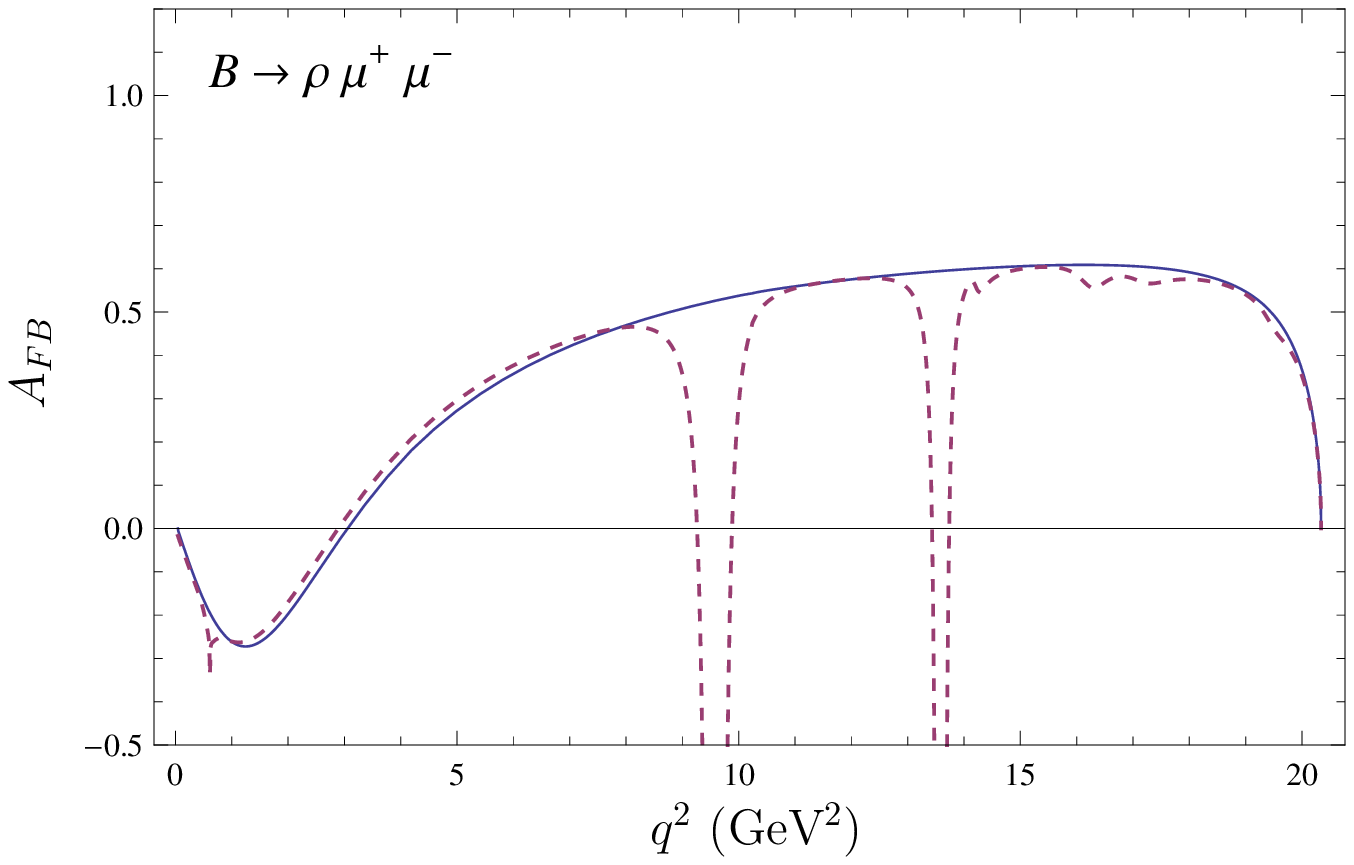}

  \caption{Same as in Fig.~\ref{fig:brbpi} but for the longitudinal polarization
    $F_L$ (left) and muon forward-backward asymmetry $A_{FB}$ (right) for the rare $B^+ \to
    \rho^+ \mu^+\mu^-$ decay.}
  \label{fig:flafd}
\end{figure}

\begin{figure}
  \centering
 \includegraphics[width=7.8cm]{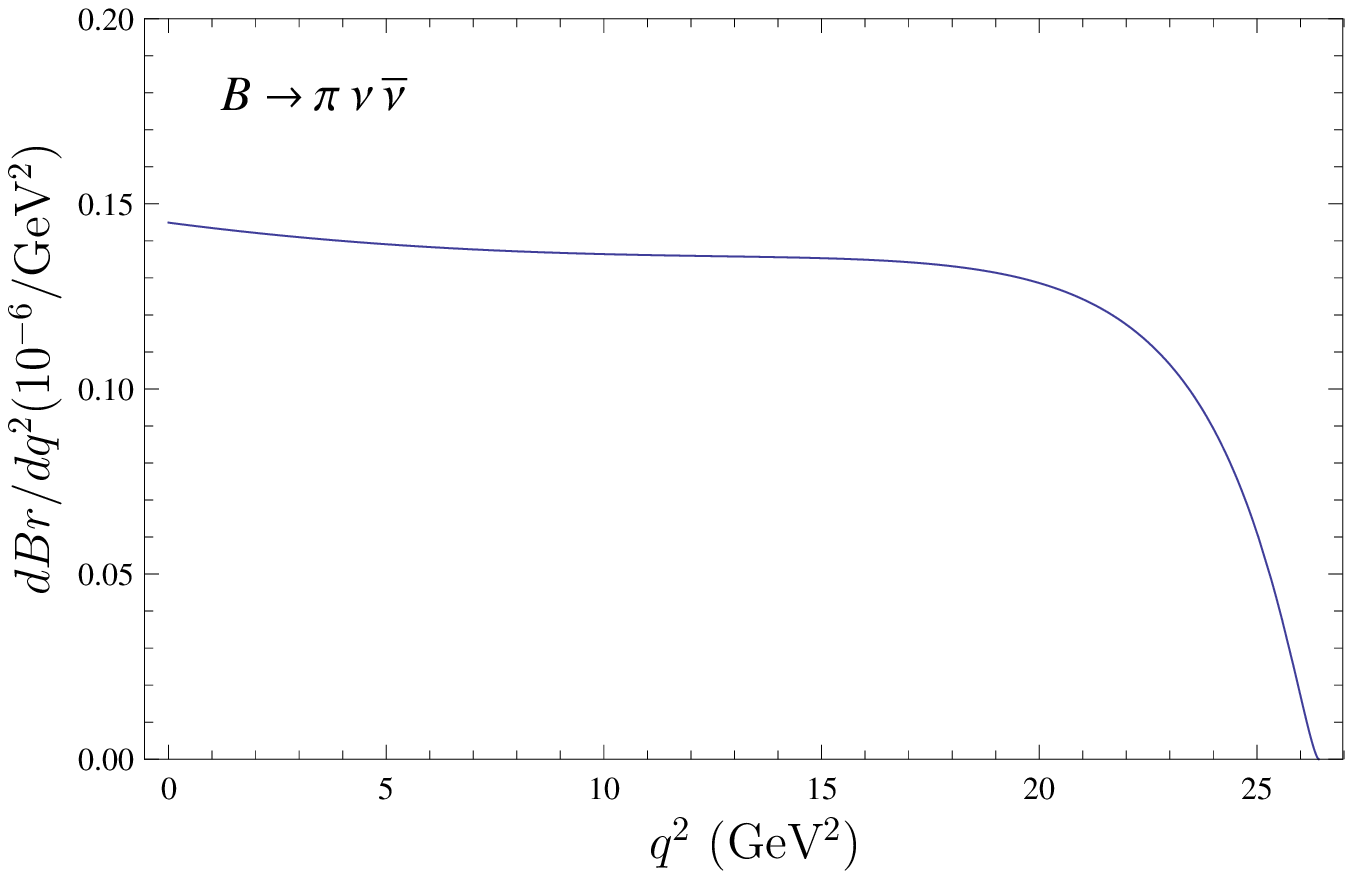} \ \
 \   \includegraphics[width=7.8cm]{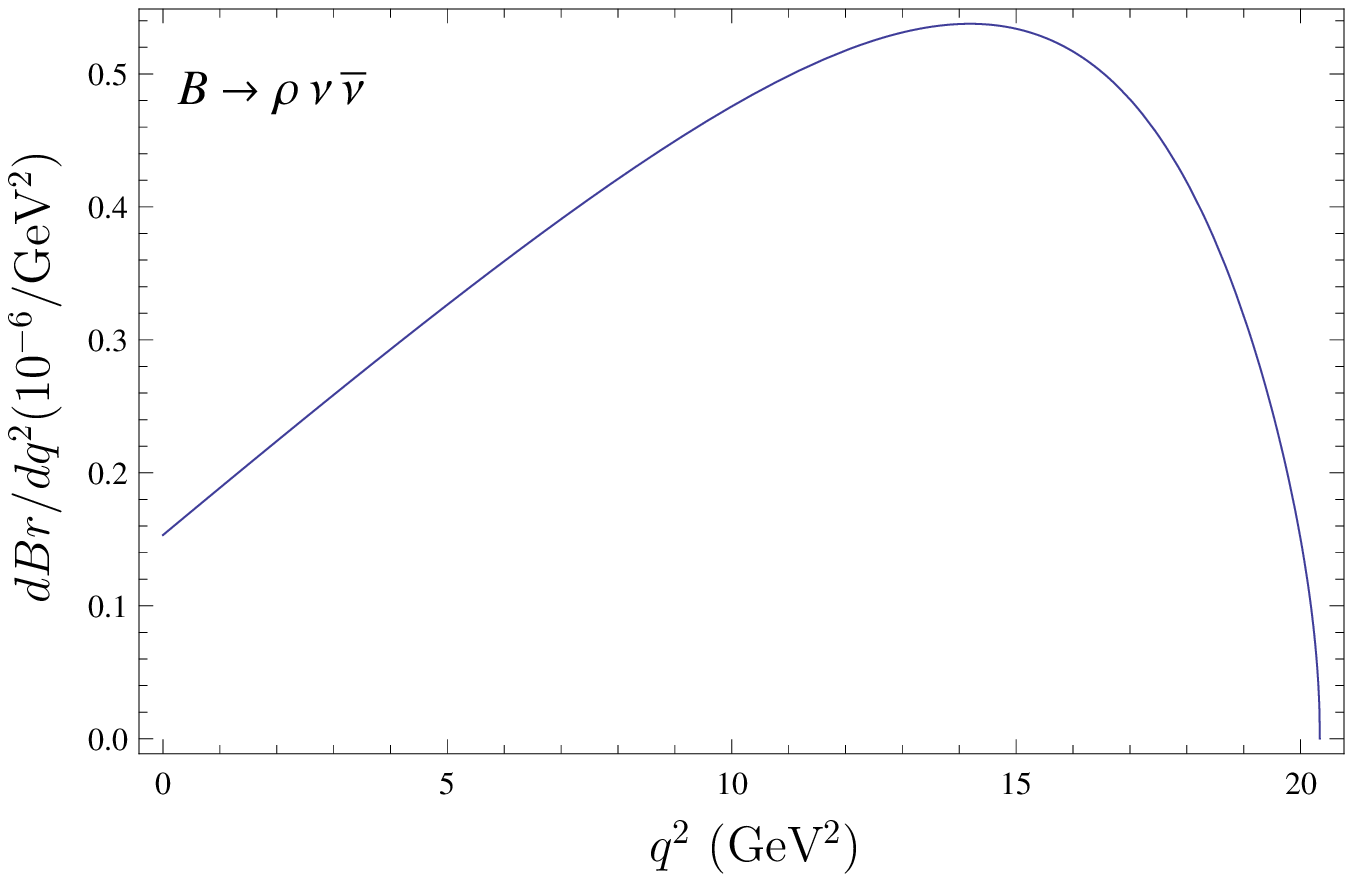}
\caption{Theoretical predictions for the differential
    branching fractions $d Br(B^+ \to \pi^+\nu\bar\nu)/d
    q^2$ (left) and $d Br(B^+ \to \rho^+\nu\bar\nu)/d
    q^2$ (right) (in $10^{-6}$). }
  \label{fig:brbnu}
\end{figure}

Substituting the form factors of the $B\to\pi$ and $B\to\rho$ weak
transitions calculated in Sec.~\ref{sec:ffbpi} in the above
expressions we get predictions for the differential decay rates, the
forward-backward asymmetry and longitudinal polarization fraction of
the vector $\rho$ meson. The obtained differential distributions for
the $B^+\to\pi^+\mu^+\mu^-(\tau^+\tau^-)$ and
$B^+\to\rho^+\mu^+\mu^-(\tau^+\tau^-)$ decays are
plotted in Figs.~\ref{fig:brbpi}--\ref{fig:brbnu}. The dashed and solid
 lines in these figures correspond to the so-called
resonant and nonresonant results which were obtained with and without
inclusion of the long-distance contributions  originating from
the $c\bar c$  and $u\bar u$ resonances [see Eq.~(\ref{eq:ybw})] in
the effective coefficient  $c_9^{\rm eff}$ (\ref{eq:c9eff}). The
regions of the highest $J/\psi$ and $\psi(2S)$  peaks are usually vetoed
in experiment in order to resolve the signal against their huge
background. Other asymmetries both time-independent and time-dependent
in these decays are discussed in detail in Ref.~\cite{bmnt}. 

In Table~\ref{brbin} we present our predictions for the differential branching
fractions of the rare semileptonic $B^+\to\pi^+ \mu^+\mu^-$ and $B^+\to\rho^+
\mu^+\mu^-$  decays integrated over several bins of $q^2$ which in
principle can be measured experimentally. In this table we also give
the recent theoretical estimates \cite{apr} for the  $B^+\to\pi^+
\mu^+\mu^-$  decay which are based on the vector weak current form
factors [$f_+(q^2)$ and $f_0(q^2)$] extracted form the combined
analysis of the available experimental data. For the tensor
current form factor [$f_T(q^2)$] lattice QCD results for the $B\to K$
transition and $SU(3)_F$-breaking Ansatz were used. We find that in
most $q^2$ bins theoretical predictions agree within error bars.    

\begin{table}
\caption{Comparison of theoretical predictions for the 
  branching fractions of the rare semileptonic $B^+\to\pi^+ \mu^+\mu^- $
  and $B^+\to\rho^+ \mu^+\mu^- $  decays
  in several bins of $q^2$ (in $10^{-8}$). }
\label{brbin}
\begin{ruledtabular}
\begin{tabular}{cccccc}
 $q^2$ bin (GeV$^2$)& \multicolumn{3}{c}{$B^+\to\pi^+ \mu^+\mu^- $} &
 \multicolumn{2}{c}{$B^+\to\rho^+ \mu^+\mu^- $} \\
\cline{2-4}  \cline{5-6}
&nonresonant&resonant& \cite{apr}&nonresonant&resonant\\
\hline
$0.05<q^2<2.00$& $0.14\pm0.02$& $0.14\pm0.02$& $0.15^{+0.03}_{-0.02}$&
$0.45\pm0.005$& $0.46\pm0.005$\\
$1.00<q^2<2.00$& $0.07\pm0.01$& $0.08\pm0.01$& $0.08^{+0.01}_{-0.01}$&
$0.10\pm0.001$& $0.11\pm0.001$\\
$2.00<q^2<4.30$& $0.17\pm0.02$& $0.20\pm0.02$& $0.19^{+0.03}_{-0.02}$& $0.23\pm0.02$&$0.26\pm0.03$\\
$4.30<q^2<8.68$ & $0.35\pm0.04$& $0.48\pm0.05$& $0.37^{+0.06}_{-0.04}$& $0.72\pm0.07$&$1.02\pm0.10$ \\
$10.09<q^2<12.86$& $0.26\pm0.03$& $0.23\pm0.03$& $0.25^{+0.04}_{-0.03}$& $0.86\pm0.08$&$0.73\pm0.07$ \\
$14.18<q^2<16.00$ & $0.16\pm0.02$& $0.14\pm0.02$& $0.15^{+0.03}_{-0.02}$& $0.55\pm0.06$&$0.48\pm0.05$ \\
$16.00<q^2<18.00$& $0.17\pm0.02$& $0.17\pm0.02$&
$0.15^{+0.03}_{-0.02}$& $0.54\pm0.06$& $0.55\pm0.06$\\
$18.00<q^2<20.34$& & && $0.36\pm0.04$& $0.32\pm0.03$\\
$18.00<q^2<22.00$& $0.32\pm0.03$& $0.28\pm0.03$&
$0.25^{+0.04}_{-0.03}$& &\\
$22.00<q^2<26.40$& $0.20\pm0.02$& $0.18\pm0.02$& $0.13^{+0.02}_{-0.02}$&& \\
\end{tabular}
\end{ruledtabular}
\end{table}

Integrating the differential branching fractions over $q^2$ we get the total rare decay branching fractions for  $B^+\to\pi^+ l^+l^-(\nu\bar \nu)$ and
$B^+\to\rho^+ l^+l^-(\nu\bar \nu)$. In Table~\ref{brbpi} we compare
our results for the nonresonant branching fractions with other
theoretical calculations. At present experimental data are only
available for the  $B^+\to\pi^+ \mu^+\mu^-$ decay \cite{lhcbBpi} which was observed recently. We see that theoretical predictions agree with each
other and the experimental value within errors.    

\begin{table}
\caption{Theoretical predictions for the nonresonant
  branching fractions of the rare semileptonic $B$ decays and
  available experimental data (in $10^{-8}$). }
\label{brbpi}
\begin{ruledtabular}
\begin{tabular}{lccccc}
 Decay& This paper &\cite{apr}& \cite{wwxy} &\cite{wx}& Experiment \cite{lhcbBpi} \\
\hline
$B^+\to \pi^+ \mu^+\mu^-$ & $2.0\pm0.2$ &$1.88^{+0.32}_{-0.21}$
&$2.03\pm0.23$&$1.95^{+1.15}_{-1.06}$ & $2.3\pm0.6\pm0.1$\\
$B^+\to \pi^+ \tau^+\tau^-$ & $0.70\pm0.07$& &&$0.60^{+0.62}_{-0.56}$&  \\
$B^+\to \pi^+ \nu\bar \nu$ & $12\pm1$ &&&$15.7^{+10.3}_{-9.5}$&\\
$B^+\to \rho^+ \mu^+\mu^-$ & $4.4\pm0.5$ & &$4.33\pm1.14$& & \\
$B^+\to \rho^+ \tau^+\tau^-$& $0.75\pm0.08$ & &&&\\ 
$B^+\to \rho^+ \nu\bar \nu$& $29\pm3$ & &&&\\ 
\end{tabular}
\end{ruledtabular}
\end{table}

\section{Conclusions}
\label{sec:concl}

The form factors parametrizing the heavy-to-light $B\to\pi$ and
$B\to\rho$ weak transition matrix elements were obtained in the
framework of the relativistic quark model. The quasipotential approach
was used to express these form factors through the overlap integrals of
the initial and final meson wave functions which are taken from the
previous calculations of meson masses. All relativistic effects,
including the wave function transformations from the rest to the moving reference
frame as well as contributions of the intermediate negative energy
states, were consistently taken into account. Our approach allowed us
to explicitly determine the form factor dependence on the momentum transfer $q^2$ in
the whole kinematical range without additional model assumptions or
extrapolations. 

First we confronted the predictions of our model for the
differential branching fractions of the semileptonic $B\to \pi l\nu_l$
and $B\to \rho l\nu_l$ decays with recent detailed experimental data
\cite{Belle1,Belle2,Babar}. Good agreement for all observables was
found. From this comparison we determined the exclusive value of the CKM matrix
element  $|V_{ub}|=(4.15\pm0.09_{\rm exp}\pm0.21_{\rm theor})\times
10^{-3}$, which is consistent with the one extracted from
the inclusive semileptonic $B\to X_ul\nu_l$ decays \cite{pdg}. 

Then we considered the rare weak $B^+\to \pi^+
l^+l^-(\nu\bar\nu)$ and $B^+\to \rho^+ l^+l^-(\nu\bar\nu)$
decays. Calculations were done both with and without account of the
long-range contributions of the heavy charmonium states and light
$\rho,\omega$ resonances. Detailed predictions for the differential
branching fractions of these decays were presented. The calculated
total branching fraction for the rare decay  $B^+\to\pi^+ \mu^+\mu^-$
agrees well with the recent measurement \cite{lhcbBpi}. The LHCb
Collaboration also measured the ratio of the $B^+\to\pi^+ \mu^+\mu^-$
and $B^+\to K^+ \mu^+\mu^-$ branching fractions to be
$0.053\pm0.014\pm0.001$. Using our prediction for the $B^+\to K^+
\mu^+\mu^-$ decay \cite{rarebk} we get the value of this
ratio equal to $0.048\pm0.005$ which agrees with the experimental one within error bars.    
The ratio of the corresponding branching fractions involving vector
$\rho$ and $K^*$ mesons is predicted to be $Br(B^+\to\rho^+ \mu^+\mu^-)/Br(B^+\to K^{*+}\mu^+\mu^-)=0.048\pm0.005$.

\acknowledgments
The authors are grateful to A. Ali, D. Ebert, D. I. Melikhov,
N. V. Nikitin, V. A. Matveev, V. I. Savrin and A. Sibidanov  
for  useful discussions.
This work was supported in part by the {\it Russian
Foundation for Basic Research} under Grant No.12-02-00053-a.

\end{document}